\documentclass[%
 reprint,
 amsmath,amssymb,
 aps,
]{revtex4-2}

\usepackage{graphicx}
\usepackage{dcolumn}
\usepackage{bm}
\usepackage{color,soul}

\begin{document}

\title{Controlled light distribution with coupled microresonator chains\\ via Kerr symmetry breaking}
\author{Alekhya \surname{Ghosh$^{1,2,\dagger}$}}
\author{Arghadeep \surname{Pal}$^{\;1,2,\dagger}$}
\author{Lewis \surname{Hill$^{\;1}$}}
\author{Graeme N \surname{Campbell$^{\;1,3}$}}
\author{Toby \surname{Bi$^{\;1,2}$}}
\author{Yaojing \surname{Zhang$^{\;1}$}}
\author{Abdullah \surname{Alabbadi}$^{\;1,2}$}
\author{Shuangyou \surname{Zhang}$^{\;1}$}
\author{Gian-Luca \surname{Oppo}$^{\;3}$}
\author{and Pascal \surname{Del'Haye$^{\;1,2,*}$}}
\affiliation{$^1$Max Planck Institute for the Science of Light, Staudtstra{\ss}e 2,
D-91058 Erlangen, Germany\\$^2$Department of Physics, Friedrich Alexander University Erlangen-Nuremberg, D-91058 Erlangen, Germany\\$^3$Department of Physics, University of Strathclyde, 107 Rottenrow, Glasgow, G4 0NG, UK}
\affiliation{$^\dagger$These authors contributed equally to this work\\$^*$Corresponding author: pascal.delhaye$@$mpl.mpg.de}

\begin{abstract}
Within optical microresonators, the Kerr interaction of photons can lead to symmetry breaking of optical modes. In a ring resonator, this leads to the interesting effect that light preferably circulates in one direction or in one polarization state. Applications of this effect range from chip-integrated optical diodes to nonlinear polarization controllers and optical gyroscopes. In this work, we study Kerr-nonlinearity-induced symmetry breaking of light states in coupled resonator optical waveguides (CROWs). We discover a new type of controllable symmetry breaking that leads to emerging patterns of dark and bright resonators within the chains. Beyond stationary symmetry broken states, we observe periodic oscillations, switching and chaotic fluctuations of circulating powers in the resonators. Our findings are of interest for controlled multiplexing of light in photonic integrated circuits, neuromorphic computing, topological photonics and soliton frequency combs in coupled resonators.
\end{abstract}

\maketitle

\section{Introduction}
When a system's physical or mathematical property remains unchanged under a certain transformation, it is said to possess symmetry. A sudden collapse of this symmetry is termed spontaneous symmetry breaking (SSB). SSB has answered pivotal questions in physics, ranging from the spontaneous breaking of gauge symmetry~\cite{RevModPhys.46.7} to more contemporary models of continuous symmetry breaking in Rydberg arrays~\cite{chen2023continuous}, the introduction of entanglement asymmetry~\cite{ares2023entanglement} and SSB in quantum phase transitions~\cite{ning2024experimental}. The applications of SSB span over a large spectrum of physics~\cite{arodz2011patterns, he2021symmetry, sym12060896, lin2019engineering}.

Kerr-ring resonators -- Kerr here referring to cubic nonlinearity ($\chi^{(3)}$) in certain materials -- have garnered interest for their capability to amass high light intensities within minuscule mode volumes, thereby enhancing the nonlinearity. These resonators have remarkable uses in optical frequency combs~\cite{del2007optical}, telecommunications~\cite{kemal2020chip}, spectroscopy~\cite{picque2019frequency}, optical clocks~\cite{papp2014microresonator}, and in sub-wavelength distance measurements~\cite{yan2024realtime}. Importantly, they also serve as experimental platforms for probing fundamental physical phenomena like SSB.

Within Kerr-resonators, SSB has been studied extensively between counter-propagating optical fields during bidirectional pumping~\cite{KAPLAN1982229, PhysRevA.32.2857,PhysRevA.98.053863,PhysRevA.101.013823,DelBino2017,PhysRevLett.118.033901,woodley2021self, campbell2022counterpropagating}. These systems have paved the way for designing optical isolators~\cite{white2023integrated}, circulators~\cite{DelBino:18} and logic gates~\cite{Moroney:20}. A second mechanism for realizing SSB originates from two co-propagating light fields with mutually orthogonal polarizations~\cite{geddes1994polarisation, PhysRevLett.122.013905, garbin2020asymmetric, huang2024coexistence, fatome2023observation}, which has led to the creation of polarization controllers~\cite{Moroney2022}, random number generators~\cite{quinn2023random}, and vectorial frequency combs~\cite{xu2021spontaneous}. Recent studies have unveiled SSB of solitons in Fabry-P\'erot Resonators~\cite{hill2023symmetry, campbell2023dark}. SSB via optomechanical effects has also been observed~\cite{Miri2017}. Recent innovations have expanded the two-field SSB phenomena to four-field SSB~\cite{hill2023multi,ghosh2023four}.

A myriad of other interesting solutions have been revealed by looking into slow-time responses, i.e., the evolutions of fields over many resonator round trips ($t_\text{r}$) in coupled cavities~\cite{cheah2023spontaneous}, photonic dimers~\cite{Miri2017, ghosh2023four}. Fast-time (time scale of a single $t_\text{r}$) responses of CROW systems~\cite{tusnin2023nonlinear} and two-dimensional microresonator arrays~\cite{mittal2021topological, flower2024observation} have also demonstrated a rich profusion of soliton dynamics. However, the slow-time response of CROW systems, rich in potential nonlinear effects, remains largely uncharted.

In this work, we conduct an in-depth study of two distinct CROW systems, and discussed the occurance of concurrent SSBs amongst different pairs of intra-resonator circulating intensities. The interplay of linear coupling and nonlinear interactions in our studied systems offers a vast parameter space to influence homogeneous responses. We demonstrate that varying input power levels lead the optical powers in the resonators to shift among different levels, exhibiting switch-like behaviors. This is promising for controllable distribution of light in photonic systems and realization of optical digital memories and computation systems. We also detect oscillations~\cite{bitha2023bifurcation, woodley2021self, Xu:22} that cause periodic interchanging of dominant field roles between distinct resonators and N-level chaotic oscillations in these systems. Precise on-chip microresonator fabrication methods~\cite{zhang2023room} will make the proposed structures soon realizable on photonic chips, thus highlighting the pertinence of the work for guiding experiments in integrated photonics.

\begin{figure}[ht]
\includegraphics[width=1\columnwidth]{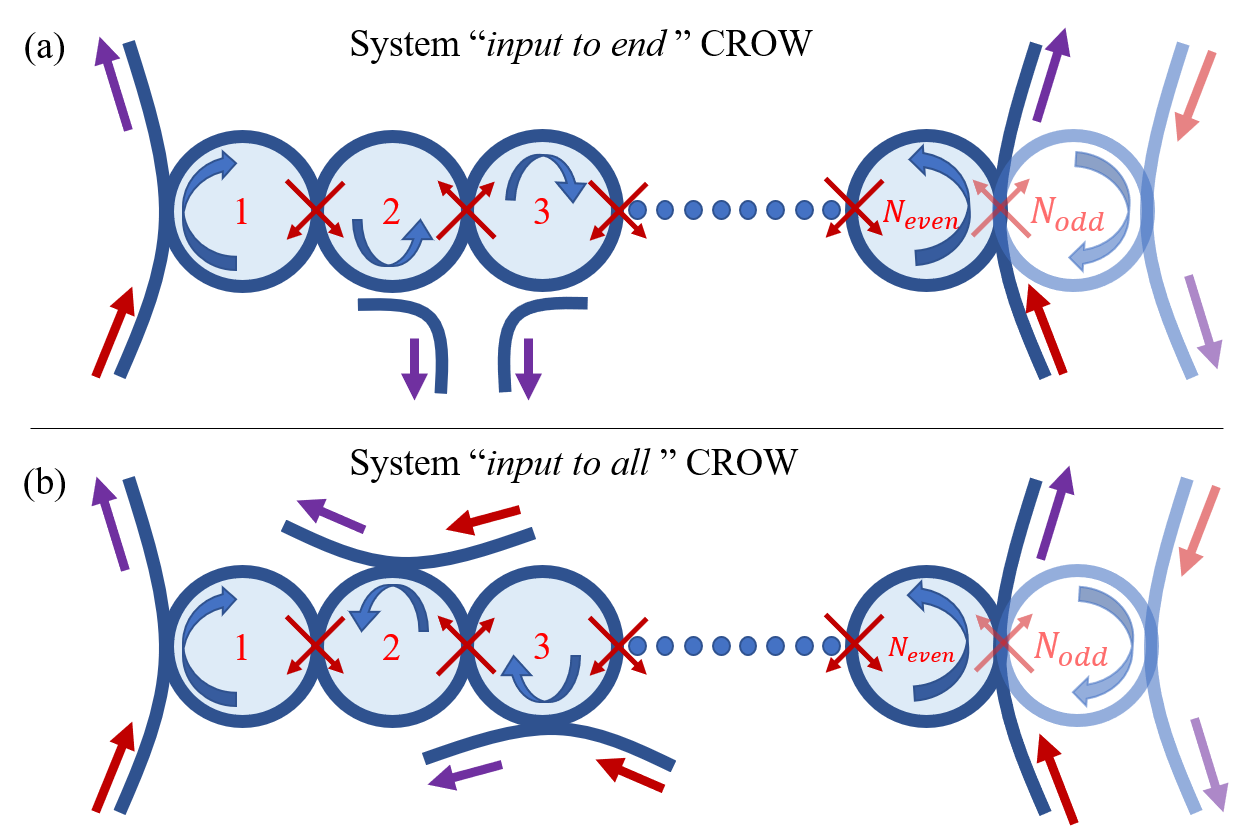}
\caption {\textit{CROW configurations.} $N$ identical Kerr ring resonators linked in sequence. Input field directions ensure that each resonator's circulating field travels only in a singular direction. Note that for systems with odd and even numbers of resonators require differing input directions for the end resonators, see faded resonators of the figure. (a) ``Input to end" CROW: Inputs are provided only to the end-resonators. (b) ``Input to all" CROW: Inputs are connected to all resonators. Input (output) directions are shown by red (purple) arrows.}
\label{schematics}
\end{figure}

\section{CROW Systems and Model}
\indent 
In our study, we consider CROW systems, as illustrated in Fig.~\ref{schematics}. The systems consist of several identical Kerr ring resonators, forming a coupled resonator chain. Here, all the light fields supplied via the input ports are assumed to be identical, i.e., they have equal intensities, frequencies, polarizations, and phases. To begin with, we consider the situation when inputs are provided to only the first and last resonators in the chain. This configuration is depicted in Fig.~\ref{schematics}(a). Afterwards, we also study the effects of facilitating inputs to all of the resonators (as shown in Fig.~\ref{schematics}(b)). Due to its more practical implementability, we first study the system with inputs only to the two resonators at the ends of the chain. The rich stationary state solutions and other homogeneous solutions of the system with more inputs are described later.
Finally, in this study, we select the directions of input light fields that result in a system without counter-propagating fields.\\
Our modelling begins with normalized coupled Lugiato-Lefever equations (LLEs)~\cite{lugiato1987spatial,tusnin2023nonlinear}. For a system encompassing $N$ coupled resonators (indexed as $n = 1...N$ for $N \geq 2$), the equations manifest as:
\begin{equation}
    \begin{split}
        \frac{\partial \psi_n}{\partial \tau} &= -\left(1 + i\zeta\right) \psi_n + i \{ (1 - \delta_{n, 1}) j\psi_{n-1} \\
        &+ (1 - \delta_{n, N}) j\psi_{n+1}\} + i|\psi_n|^2\psi_n + \alpha_n f,
    \end{split}
    \label{LLEquations}
\end{equation}

\noindent where $\psi_n = \sqrt{2g_0/\kappa}A_n$ is the normalized optical field envelope in the $n^\text{th}$ resonator, and within which $A_n$ is the unnormalized field envelope, $g_0$ is the Kerr gain, and $\kappa = \kappa_l + \kappa_e$ is the total cavity losses, with internal losses $\kappa_l$ and external losses $\kappa_e$. The normalized cavity detuning is given as $\zeta = 2\Delta/\kappa$, where the unormalized cavity detuning is given as $\Delta = \omega_0 - \omega_{\text{res}}$ (the difference between the input laser frequency ($\omega_0$) and the closest cavity resonance frequency ($\omega_{\text{res}}$)). The  terms within the curly brackets account for the inter-resonator couplings, where $\delta_{p,q}$ is the Kronecker Delta function and $j = 2J/\kappa$ is the normalized inter-resonator coupling rate, with $J$ being the unnormalized coupling rate. Input to the resonators is given as $f = \sqrt{8\kappa_{e}g_0/\kappa^3} s_{\text{in}} e^{i\phi_{\text{in}}}$, where, $s_{\text{in}}$ and $\phi_{\text{in}}$ are respectively the input pump amplitude and the corresponding phase. In our simulations we assume $phi_\text{in}$ to be constant and set it to $phi_\text{in} = 0$ for convenience. The term $\alpha_n = 1$ when input is provided to the $n^\text{th}$ resonator and $\alpha_n = 0$ otherwise. The normalized intracavity intensity in the $n^\text{th}$ resonator is given by $\Psi_n = |\psi_n|^2$ and the input power by $F = |f|^2$. We have neglected dispersion in the systems. The second term (the term within the curly brackets) on the RHS of Eq.~\eqref{LLEquations} describes the fact that all the resonators in the chain are coupled to the previous and the next resonator in the line except for the two end resonators, each of which is just connected to one adjacent resonator. The third term of Eq.~\eqref{LLEquations} is the self-phase modulation term, which accounts for the nonlinear effect of a field on itself. The last term on the RHS of Eq.~\eqref{LLEquations} represents input from outside the system. Since all the ring resonators in both cases are identical, parameters, such as the cavity detunings and the Kerr nonlinear gains, $g_0$, are the same for all resonators.

\begin{figure*}
\includegraphics[width=1.5\columnwidth]{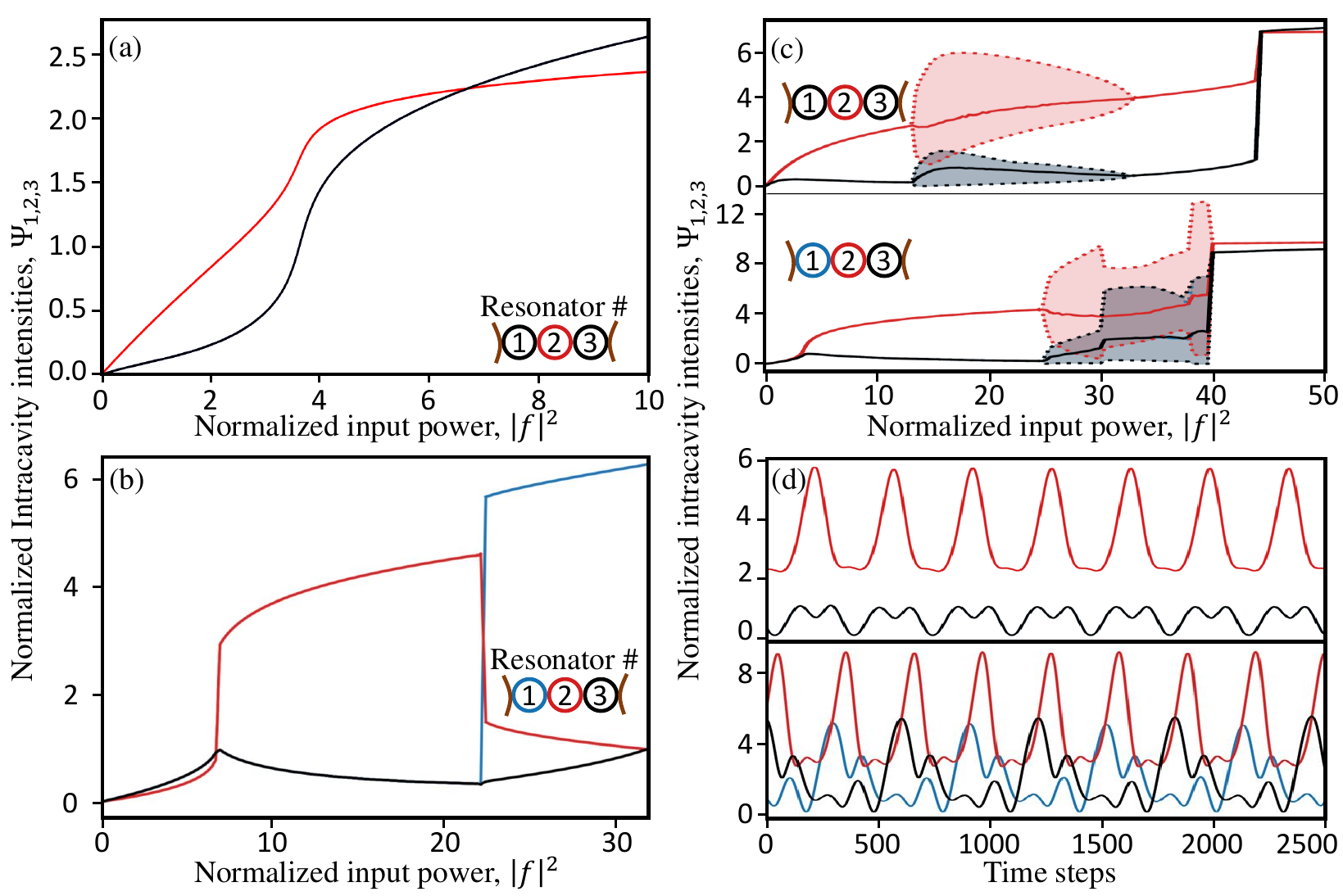}
\caption {\textit{Evolutions of optical intensities in $N=3$ ``input to end" CROW system.} 
Panels (a)-(c) show the evolutions of the field intensities in different resonators as a function of input power. The field intensities in the end resonators are depicted in blue and black, whereas, the field intensity in the middle resonator is depicted in red. For (a), $\zeta = 0.5$, $j = 1$. A field intensity crossing point appears in (a). Before the crossing point, the field intensity in the middle resonator is higher than the fields in the end resonators and beyond this point, the end resonator fields become more intense than the middle resonator field. Panel (b) depicts spontaneous symmetry breaking of the end resonator light field intensities for $\zeta = 5$, $j = 2$. Symmetry unbroken and symmetry broken oscillations are depicted in the upper ($\zeta = 3$, $j = 2$) and lower ($\zeta = 5$, $j = 2.5$) panels of (c) respectively. In this and all successive figures, the dotted lines stand for the positions of the maxima and the minima of the oscillations, and the shaded regions in between highlight the span of the oscillations. Upper and lower panels of (d) are examples of symmetry unbroken and complete symmetry broken oscillations respectively. Used parameters: $|f|^2 = 22,\zeta = 3,j = 2$ (d - upper panel), $|f|^2 = 37.79,\zeta = 5,j = 2.5$ (d - lower panel). Time step for integration is $0.005$.
}
\label{IECROW_powScans_3}
\end{figure*}

\section{$\mathbf{N=3}$ CROW systems}
Our present analysis addresses CROWs with three or more resonators, i.e. $N \geq 3$. The homogeneous states of the systems are obtained by numerically evaluating Eq.~\eqref{LLEquations} for a variety of initial conditions and over sufficient evolution times. The stationary states of the systems, which form a subset of the homogeneous states, where the circulating fields remain unchanged over time, can be obtained by setting $\partial \psi_n/ \partial \tau = 0$. For an $N=3$ system, analytical solutions for the stationary states can be derived (See Appendix~\ref{A} for more details), but for $N>3$ systems, obtaining analytical solutions is difficult.
Since for $N \geq 3$, the two end resonators are connected to only one resonator while all other resonators in the CROW system are connected to two neighbouring resonators, there is an inherent asymmetry in the system. The circulating field intensity in each resonator $n$ is asymmetric to that of resonator $m$ for $m \neq N-n+1$. This asymmetry comes from the linear coupling terms of Eq.~\eqref{LLEquations}. Consequently, resonator $n$ is symmetrical to only resonator $N-n+1$ in terms of coupling arrangements. In other words, the field intensities are symmetric around the center of the chain.

Figure~\ref{IECROW_powScans_3} shows different kinds of optical intensity distributions that can be observed in $N=3$ CROW systems. In Fig.~\ref{IECROW_powScans_3}(a), it can be observed that for lower input powers, intracavity field intensities in the end resonators behave symmetrically, but the field intensity in the middle resonator is more than that of the individual end resonators. However, for a certain value of input power, the field intensities in the end resonators cross the field intensity in the middle resonator, and grow steadily after that, with the middle resonator's light intensity remaining almost constant. At the crossing point, there is a momentary occurrence of complete symmetry, where all three resonators have the same field intensities. Around the crossing point, the relative distribution of optical intensities between the middle resonator and the two end resonators can be tuned in a very controlled manner by changing the input power (discussed in Appendix~\ref{B}). Figure~\ref{IECROW_powScans_3}(b) reveals a spontaneous symmetry breaking of the circulating optical intensities in the end resonators. This adds to the various optical field distribution mechanisms that can be achieved in the CROW systems. It is important to note that the SSB depicted in Fig.~\ref{IECROW_powScans_3}(b) is a novel mechanism, quite different from the usual SSB observed in Kerr-resonators~\cite{PhysRevA.98.053863, DelBino2017, PhysRevLett.122.013905, ghosh2023four}. The detailed description of the emergence of this SSB phenomenon is discussed in the next section. Following the trajectory of the field intensity in the middle resonator (in red) in Fig.~\ref{IECROW_powScans_3}(b), an interesting characteristic can be observed. The field intensity remains low for low input powers, jumps to a high value after a certain input power and finally comes back to a low value in the SSB region. This effectively allows the system to be used as an all-optical switch with certain low and high cut-off powers.  Apart from the stationary states, numerical simulations of Eq.~\eqref{LLEquations} also reveal other homogeneous solutions, such as slow-time oscillations inside the resonators of the chain. Such Kerr induced oscillations of the field intensities have been observed here with and without the occurrence of SSB in the system, as depicted in the upper and lower panels of Fig.~\ref{IECROW_powScans_3}c respectively. Oscillations of all field intensities are observed in the upper and lower panels of Fig.~\ref{IECROW_powScans_3}d respectively. In the upper panel, the middle resonator field intensity dominates over the end resonator field intensities, which always oscillate in phase. On the other hand, the lower panel shows a perfect periodic switching of the field intensities in the end resonators, where each of the three field intensities becomes dominant over the other two at certain instances. All of the observed phenomena pave the way for the $N=3$ system to become an efficient option for optical field routing in integrated systems.

\begin{figure*}
\includegraphics[width=1\textwidth]{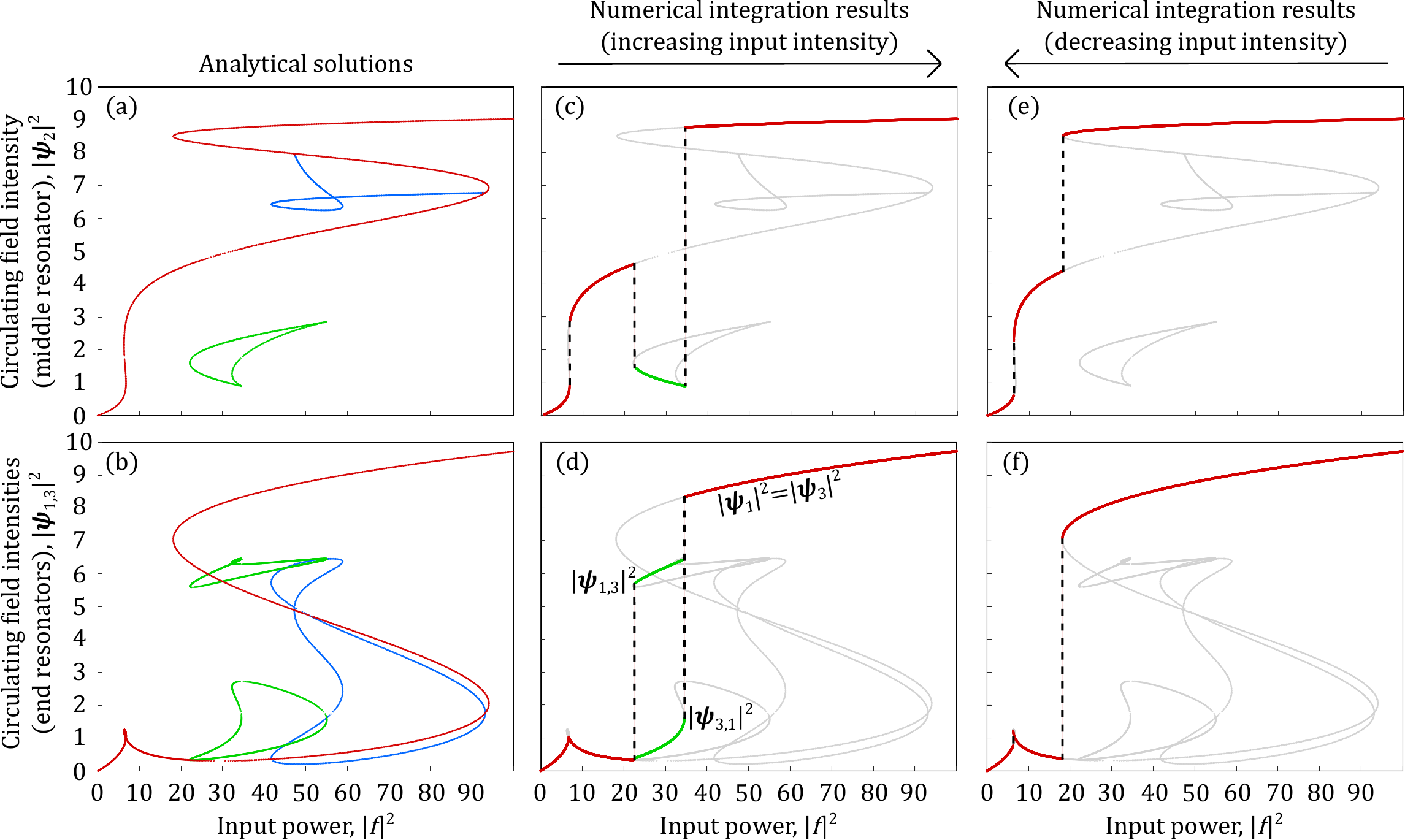}
\caption {\textit{Circulating field intensities, $\Psi_n (= |\psi_n|^2)   $, against input power, $|f|^2$, for an $N = 3$ ``input to end" CROW System.}
For detuning $\zeta=5$ and inter-resonator coupling $j=2$, we present in panels (a, b) the analytical solutions to Eq.~\eqref{LLEquations} for the fields circulating the middle $\Psi_2$ and end resonators $\Psi_{1,3}$ respectively. In panels (c, d) and panels (e, f), respectively, we display the results of numerical integrations of Eq.~\eqref{LLEquations} for stepwise increasing, and stepwise decreasing values of the input field intensity. In all panels, different “relationships” of solutions are coloured accordingly for visual benefit, by this we mean that when fields $\Psi_{1,3}$ are on the green solution line, this means that $\Psi_2$ is also on its own respective green solution line. These results are discussed thoroughly in the main text, but we highlight the possibility of end-resonator-symmetric solutions (red) and two distinct end-resonator-symmetry-broken solutions (green and blue).
}
\label{analytical}
\end{figure*}
\section{Analytical solution for $\mathbf{N=3}$ CROW system}
Input power, $|f|^2$, scans for an $N=3$ CROW, when inputs are provided to only end resonators, are presented in Fig.~\ref{analytical}. In Fig.~\ref{analytical}(a,b) the analytical solutions to Eq.~\eqref{LLEquations} are displayed. One may find details on the analytical solutions to Eq.~\eqref{LLEquations} in the Appendix~\ref{A}. Panels (a,b) reveal surprisingly rich and interesting dynamics for such a simple system. As mentioned earlier, for nonzero input powers, $\Psi_2 \neq \Psi_{1,3}$ due to coupling conditions. The difference in coupling causes persistent differences in the evolutions of the fields circulating the resonators, quite evident from the vastly different solution curves of panels (a) and (b).
\begin{figure}[t]
\includegraphics[width=1\columnwidth]{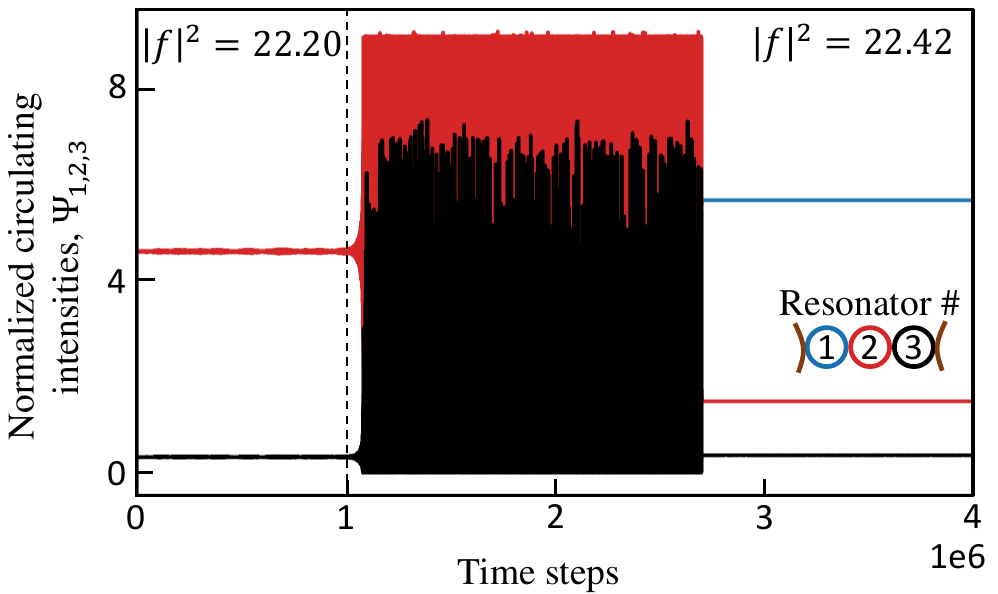}
\caption {\textit{Occurance of SSB in N=3 CROW system.}
The left side of the dashed line corresponds to a scan with an input power of $22.20$, whereas, the right side of the scan corresponds to a scan with an input power of $22.42$. It can be seen that starting from a stable ``black-blue symmetric" state, the system after the change in input power, goes to a ``black-blue symmetry broken" state after a small oscillatory evolution. The origin of the oscillations can be described from the linear stability analysis presented in Appendix~\ref{C} and~\ref{D}. The blue and black lines represent the circulating field intensities in the end resonators, whereas, the red line depicts the field in the middle resonator. The vertical dashed black line marks the point where the input power is increased. In each case, the detuning is $5$. Time step for integration is $0.005$.
}
\label{SSB_step}
\end{figure}
The analytical solutions of the end resonators, Fig.~\ref{analytical}(b), reveal not one but two distinct sets of asymmetric solutions occurring for the end resonators. One set (blue) arises in a manner similar to that of previous studies – a pitchfork bifurcation emanating from the (red) symmetric solution line. The more surprising part is the second asymmetric solution set (green), which does not originate from a pitchfork bifurcation of the symmetric line. Indeed, it does not originate from the symmetric solution line at all. The green solution sets form entirely isolated solution “bubbles”, an isolation seen most prominently in Fig.~\ref{analytical}(a). Isolated sets of asymmetric solutions have not been seen in any past works, to our knowledge, with the exception of a significantly different setup with unbalanced input conditions~\cite{garbin2020asymmetric}. A justified question to ask is the following; as interesting, perhaps, as the discovery of a novel SSB origin is, if these solutions are isolated, does this not imply that they are unreachable under experimental conditions – and hence entirely useless? We report that this is, highly surprisingly, not actually the case.
Panels (c-f) of Fig.~\ref{analytical} show, as a counterpart of the discussed analytical results, the results of the numerical integration of Eq.\eqref{LLEquations} via standard Runge-Kutta methods (Fig.~\ref{analytical}(c) and (d) are extended versions of Fig.~\ref{IECROW_powScans_3}(b)). In panels (c,d) the input power is stepwise increased following a suitable system relaxation time, while in panels (e,f) the input power is similarly stepwise decreased. Unlike the analytics, which provide the full solution sets, these scans predict the real-world behaviors and evolutions of the circulating fields under experimental conditions. From panel (d), we see that in the input-increasing-scan, just after $|f|^2=20$, the field intensities of the end resonators suddenly jump away from the red symmetric solution line and begin evolving, instead, along the green, isolated, asymmetric solution line. This is accompanied by a substantial drop in the middle resonator power, as seen in panel (c). The obvious question is how? How does the system find the isolated set? The answer lies in the stability of Eq.~\eqref{LLEquations}. Performing a linear stability analysis, in Appendix~\ref{C}, we find that, at the point that the isolated asymmetric solutions occur, the symmetric solution line experiences a Hopf bifurcation (Appendix~\ref{D}) leading to system oscillations with wide-ranging intensity changes. These oscillations allow for the system to eventually find, and settle on, attractive and stable, but isolated, asymmetric solutions. This process is shown in Fig.~\ref{SSB_step}. As the increasing-input-scan continues further an optical bistability in the asymmetric states leads to the system losing stability once again and proceeds, this time, to settle on the stable upper branch of the original symmetric solution line.
Owed to the strong stability of this upper branch, we find that a reverse scan in this case reveals none of the asymmetric solutions of the forward scan, only bistability jumps.
Fig.~\ref{analytical} has revealed that even for low values of $N$, Eq.~\eqref{LLEquations} describes a system capable of extremely intricate dynamics.

\begin{figure*}[t]
\includegraphics[width=1\textwidth]{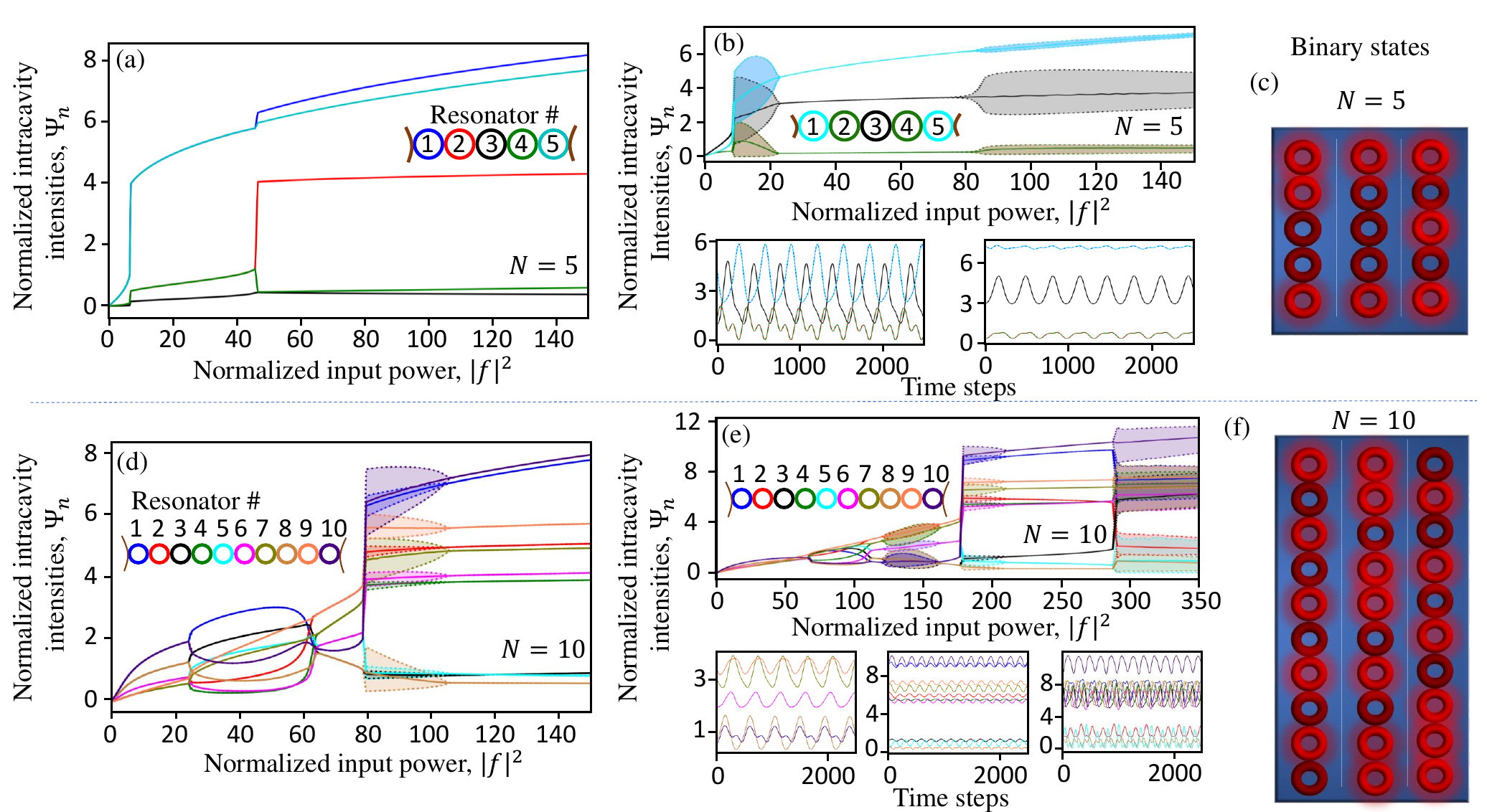}
\caption {\textit{Evolutions of optical intensities in $N=5$ (a-c) and $N=10$ (d-f) ``input to end" CROW systems.}
Panels (a),(b - upper panel) show the evolutions of the field intensities in different resonators as a function of input power. The end resonator field intensities are depicted in cyan and blue. In (a), the end resonator field intensities and the neighboring resonator's field intensities (depicted in red and green) display spontaneous symmetry breakings. In (b), the field intensities within coupling-wise symmetric resonators always remain symmetric and they display oscillations. Overlapping (lower-left panel) and non-overlapping (lower-right panel) oscillations are observed. In (c), we demonstrate three of the possible light intensity distribution conditions in the resonators with bright-red (dark-red) referring to bright (dark) resonators (left configuration: $\zeta = 3.62, J = 1, |f|^2 = 59.18$, middle configuration: $\zeta = 3.62, J = 1, |f|^2 = 37.88$ and right configuration: $\zeta = 3, J = 2, |f|^2 = 112.99$). Input power scans for $N=10$ CROW systems are depicted in (d) and (e-upper panel). All the symmetric field pairs undergo SSBs (in intensity) in (d) and upper panel of (e), followed by bistability jumps and symmetry broken oscillations. Moreover, symmetry unbroken oscillations are observed in the upper panel of (e). In the lower panel of (e), three examples of oscillations of field intensities in time are depicted from the different oscillatory regions as shown in the corresponding upper panel. Panel (f) presents three possible light distribution conditions (left configuration: $\zeta = 1.5, J = 4, |f|^2 = 37.88$, middle configuration: $\zeta = 1, J = 6, |f|^2 = 270.67$ and right configuration: $\zeta = 1, J = 6, |f|^2 = 88.38$). Used parameters: (a) $\zeta=3.62$ and $j=1$, (b) $\zeta=3$ and $j=2$, (d) $\zeta=1.5$ and $j=4$, and (e) $\zeta=1$ and $j=6$. Time step for integration is $0.005$.
}
\label{IECROW_powScans_5_10}
\end{figure*}
\section{CROW systems with $\mathbf{N>3}$}
To reveal the full potential of the CROW systems, we continue to study the homogeneous solutions in $N>3$ systems. Figure~\ref{IECROW_powScans_5_10} shows SSB phenomena occurring in the CROW systems with $N = 5$ and $N = 10$, with inputs provided only to the end resonators. In both systems, SSB bifurcations can be observed between different symmetric field pairs. 

For lower input powers in the $N=5$ system, $|\psi_1|^2=|\psi_5|^2 \neq |\psi_2|^2 = |\psi_4|^2 \neq |\psi_3|^2$ \& $|\psi_3|^2 \neq |\psi_1|^2 $. In Fig.~\ref{IECROW_powScans_5_10}(a), where $\zeta=3.5$ and $j=1$, the field intensity in the middle resonator (in black, index $3$) is suppressed greatly for all input power values. The intensities of the fields within the end resonators (index $1$ and $5$ - depicted in blue and cyan) and the set of resonators coupled to the end resonators (index $2$ and $4$ - depicted in red and green) display spontaneous symmetry breakings. The system also shows bistability jumps. In the upper panel of Fig.~\ref{IECROW_powScans_5_10}(b), we can observe two isolated regions with symmetry unbroken oscillations in the system. The lower panel shows oscillations of circulating field intensities in time corresponding to the two different regions. Fig.~\ref{IECROW_powScans_5_10}(c) depicts some of the possible steady state light field intensity distributions among different resonators for the $N=5$ CROW system. Here, the state of a resonator is assumed to be bright if the circulating field intensity within the resonator is more than the average of all the resonator's field intensities in the respective CROW arrangement for a particular combination of system parameters and input powers.

Fig.~\ref{IECROW_powScans_5_10}(d-e) depicts the input power scan of the $N=10$ system for two different sets of parameters. In Fig.~\ref{IECROW_powScans_5_10}(d), independent SSBs of all field intensities within resonators with symmetric coupling conditions are observed, with SSB bubbles crossing each other. The initial SSB region is followed by a symmetry restored region, which then is followed by regions of oscillating full asymmetric solutions, and subsequently non-oscillating full asymmetric solutions. Fig.~\ref{IECROW_powScans_5_10}(e) depicts SSB of all coupling-wise symmetric pairs, symmetry restored regions with and without oscillations, and bistability jumps of the circulating field intensities leading to a second region of full asymmetry. This full-asymmetric region also displays oscillations for certain ranges of input powers with and without overlaps. The lower panels of Fig.~\ref{IECROW_powScans_5_10}(e) show examples of field intensity oscillations in time corresponding to three different oscillatory regions. 

Fig.~\ref{IECROW_powScans_5_10}(f) shows some of the possible bright-dark conditions achievable in $N=10$ CROW systems. These demonstrate the power distribution capabilities of the CROW systems. Different arrangements of bright-dark resonators, achievable via tuning the input power can be used as different binary states. Therefore, the system can be used as an optical analog-to-digital converter where an analog optical input is transformed into a digital binary bit-string and multi-bit logical operations can be performed on them. Moreover, in Ref.~\cite{tusnin2023nonlinear}, the authors have discussed the possibilities of having different dynamical fast-time solutions in the systems. These solutions, e.g., solitons, depend on the interplay of the dispersion profiles and nonlinear gains in the systems. The power redistribution that can be achieved in CROW systems can significantly change the intensity-dependent nonlinear gains in the resonators, affecting the fast-time dynamics. Therefore, by exploiting the interactions of the Kerr effect and coupling between resonators one can gain control over the fast-time dynamics in these systems. 
\begin{figure}[t]
\includegraphics[width=1\columnwidth]{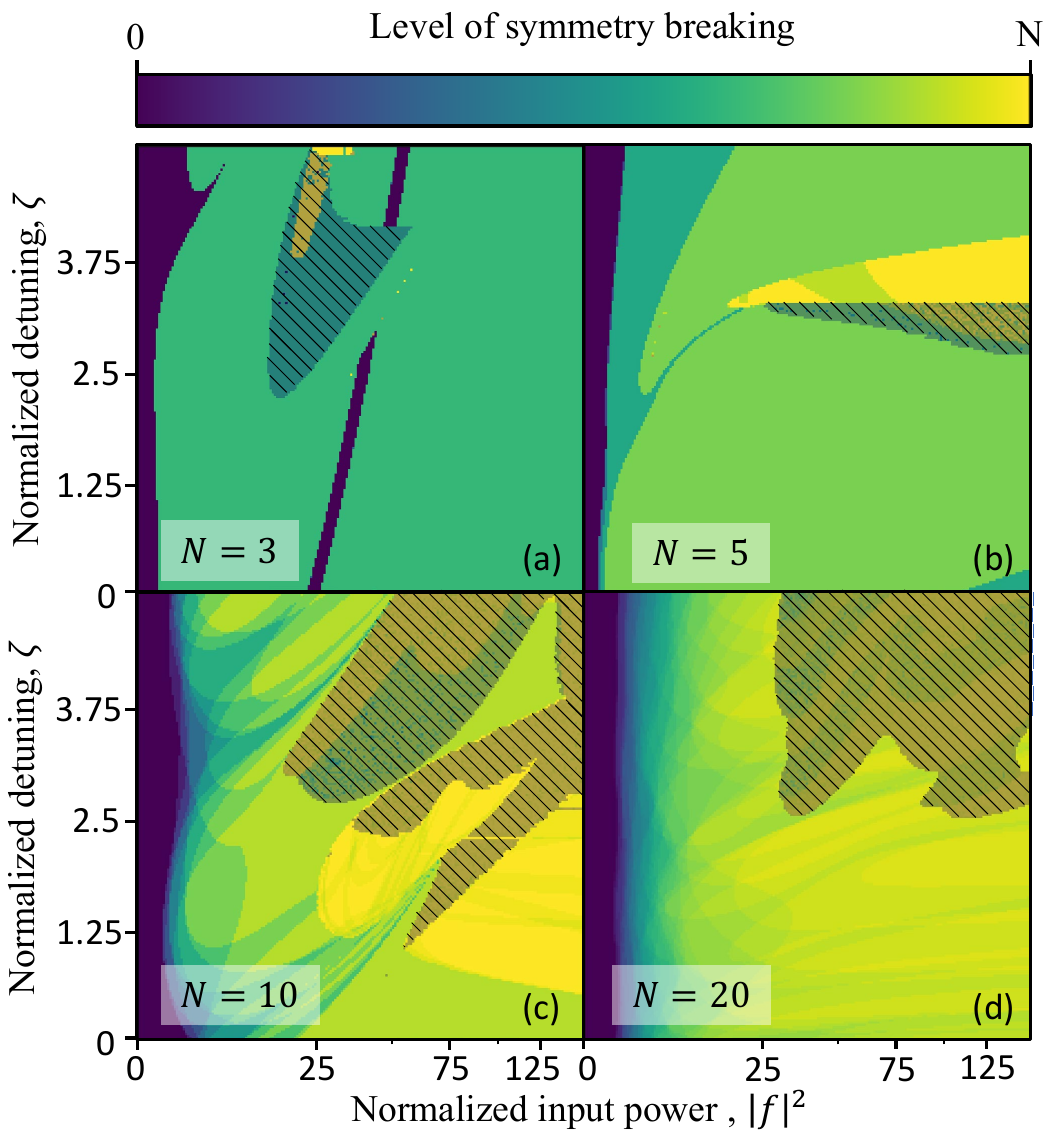}
\caption {\textit{Colormap of different symmetry breaking conditions as function of input power and detuning for $N=3$ (a), $N=5$ (b), $N=10$ (c) and $N=20$ (d) ``input to end" CROW systems}. 
Purple regions stand for symmetry unbroken case and the yellow stands for completely asymmetric case. The colors in between (different shades of green) stand for different levels of symmetry breakings in the systems. The watershed portions with stripes in each panel display the regions of oscillations. In all cases, the scans are done from lower to higher input powers for certain detuning values. In other words, Eq.~\eqref{LLEquations} is scanned for all detuning values starting from zero input power. For each input power, the initial values of the field amplitudes are selected to be the steady state value of the last step (smaller input power). The scans for all detunings are done in parallel. For the scans $j = 2 \text{(a), }1 \text{(b), }3.5 \text{(c), and }3 \text{(d).}$
}
\label{param_scans_IECROW}
\end{figure}

Motivated by the fact that a rich variety of SSBs can be observed in the CROW system, we performed input power-detuning scans for $N = 3,5,10,20$ and the corresponding results are depicted in Fig.~\ref{param_scans_IECROW}. All the scans are performed for increasing input power. The scans demonstrate different thresholded symmetry breaking conditions and oscillations of the circulating field intensities through different colored regions. Two fields are considered to be symmetric if the difference of their normalized intensities lies within an upper limit (here we choose 0.05). These scans can be used to allocate different amounts of homogeneous power in different resonators.

\begin{figure*}[t]
\includegraphics[width=1\textwidth]{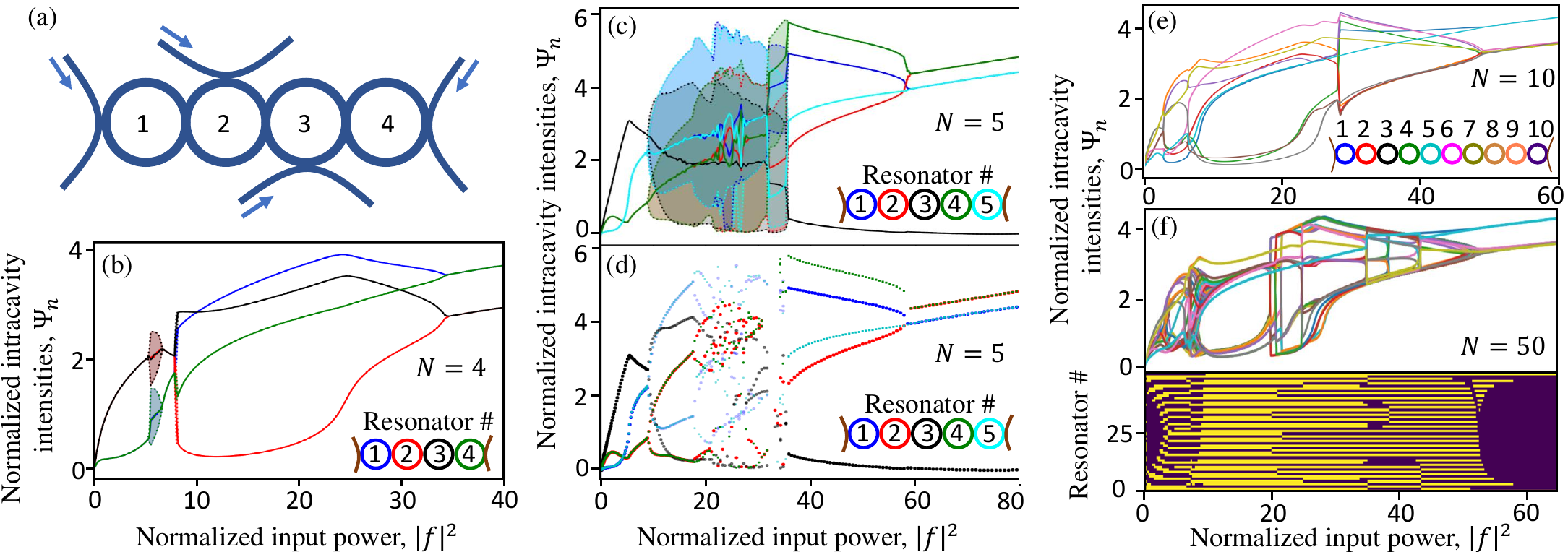}
\caption {\textit{Evolutions of optical intensities in ``input to all" CROW systems for $N=4$ (b), $N=5$ (c), $N=10$ (e) and $N=50$ (f).}
Panel (a) depicts the schematics for a system with $N=4$. Panel (b) presents the evolutions of the field intensities in an $N = 4$ CROW system with increasing input power for $\zeta = 1.5$ and $j= 1$. Oscillations, bistability jumps and SSBs are observed sequentially in the system with increasing input power. Panel (c) shows the evolutions of the field intensities in an $N = 5$ CROW system with increasing input power for $\zeta = 2.5$ and $j= 2$. Panel (d) displays the Poincar\'e section plot of (c). In the region where dots of all colors look randomly scattered, chaos is observed. Panel (e) and the upper panel of (f) present evolutions of the field intensities for $N = 10$ ($\zeta = 1.5$, $j=1$) and $N = 50$ ($\zeta = 1.5$, $j = 1$) CROW systems respectively. In both cases, the fields within coupling-wise symmetric resonators break their symmetry with increasing input power, afterwards, they form two bunches, one at high power and one at low power. Thereafter in panel (e), by a bistability jump, the field intensities form two inverse bifurcation structures. In the upper panel of (f), several bistability jumps can be observed, where different field intensities group and regroup at different jumps. Finally, two inverse bifurcation structures form. The lower panel of (f) demonstrates the bright (yellow)-dark (purple) conditions of different resonators for different input powers.
}
\label{IACROW}
\end{figure*}
\section{``input to all" CROW systems}
In this section of the paper, we aim to study the CROW configurations where all the resonators are provided with input fields. Since in these systems, all the resonators in the chain have access to input power (as shown in Fig.~\ref{IACROW}(a)), richer nonlinear effects are expected to be observed. The high degree of controllability of fields, due to inputs to all the resonators, makes this section important, especially for the experimentalists working on integrated coupled resonator systems.

For $N = 4$, the full asymmetry of circulating field intensities in different resonators is depicted in Fig.~\ref{IACROW}(b) for a wide range of input power values. The end resonator field intensities (in blue and green) initially remain symmetric and have less power than the middle resonators (index $2$ and $3$, shown in black and red). With increasing input power, before the SSB region, a region of oscillation appears. The near-switching behaviors of the field intensities are depicted in the Appendix~\ref{E}. It is important to note that in the full asymmetric region, the crossing of the various fields' intensities leads to two very localized regions of three-level asymmetry. These particular crossings are intriguing since despite the different coupling conditions for the end and middle resonators, at some distinct input power levels one of the middle resonators becomes symmetric to one of the end resonators. The two SSB bubbles close by forming inverse bifurcation structures after a certain input power leading to two separated pairs of circulating field intensities. Here, each pair contains the fields within resonators that have symmetric coupling conditions, as observed for lower input powers. It can be observed that the dominance order of intensities of the pairs switches between before and after the SSB region.

With increasing $N$, more and more interesting nonlinear phenomena are observed. For an $N = 5$ system, Fig.~\ref{IACROW}(c) shows the input power-dependent redistribution of relative optical intensities among different resonators. The intensity of the middle resonator field (index = $3$, in black) gradually decreases from being higher than any other fields at small input powers to being extremely suppressed after the complete symmetry breaking of all the fields. Figure~\ref{IACROW}(d) gives an insight into the oscillations in the system which is portrayed via the Poincar\'e section plot, where the maxima and minima of the temporal oscillations of the respective field intensities are presented by dots for each input power. At the beginning of the oscillatory regime, the maxima and minima of the initially symmetric field intensities overlap completely. Therefore three regions of oscillations appear, the overlap regions of which gradually increase with input power. After the region of periodic oscillations, the system drives into the region of chaos with increasing input power, where all the field intensities oscillate asymmetrically. This symmetry broken chaotic oscillation region ends in symmetry broken stationary states. The full asymmetry closes with $2$ inverse bifurcation structures. The SSB enforces high suppression of circulating power in the middle resonator, whereas, the outer resonators always have higher circulating field intensities. A detailed overview of the oscillations for the $N = 5$ system is given in Appendix~\ref{E}.

For $N = 10$, as shown in Fig.~\ref{IACROW}(e), we again observe an SSB region, however, the SSB bubbles here twist and cross each other to provide different levels of symmetry breakings. At the end, following bistability jumps, the field intensities in the resonators form two inverse bifurcation structures, one by the fields within the end resonators, and one by all other fields. It is noteworthy that all the inner resonators at this point behave almost symmetrically. For $N = 50$, depicted in the upper panel of Fig. 6(f), SSB bubble crossings generate a much more complex scenario with many possible levels of SSB. However, a noticeable phenomenon in this case is the group formation of the field intensities within the SSB bubble, where two pairs of fields with little differences in intensities emerge. These intermediate pairs split up with multiple bistability jumps and a series of regroupings occur. Finally, the field intensities merge into two symmetric pairs through two inverse bifurcation structures. For the first time, it has been observed that the field intensities jump between different levels of circulating power, forming a unique cage-like diagram in the regrouping section. Even if various resonators in the system have coupling-wise asymmetry, it is seen in both Fig.~\ref{IACROW}(e) and (f - upper panel), after the inverse bifurcations only two bunches sustain, one by the fields within the end resonators, and one by all other fields. All the inner resonators at this point behave almost symmetrically. These SSB-induced intra-cavity field distributions in coupled resonator systems are not only intriguing for generating multi-logic levels with higher functionalities in all-optical devices but also give an idea for observing various soliton dynamics at different input power levels.

In the lower panel of Fig.~\ref{IACROW}(f), we demonstrate the corresponding dark-bright conditions of different resonators as a function of input power. As mentioned earlier, a resonator is considered to be bright (shown in yellow) if the field intensity within that resonator is more than the average of all the field intensities of all the resonators. Otherwise, it is considered to be dark (shown in purple). The dark-bright condition plot shows that these configurations of CROW systems have great potential in all-optical computing. The input-output access to all the resonators makes it ideal for loading and unloading a bit-stream of data in a digital computing platform, with the ability to perform logical operations via the control of optical field distributions within the resonators.

\section{Discussions and outlook}
\indent To summarise, a theoretical framework has been developed to examine different states of light in CROW systems. In CROW systems with inputs to the end resonators, we can observe a plethora of different symmetry breaking phenomena. The spontaneous symmetry breaking causes different light field intensities in different resonators. At low input powers, mirror symmetric pairs of resonators within the chain experience equal circulating powers. When the input power is increased, more complex light distributions within the resonator chain emerge, corresponding to multiple concurrent spontaneous symmetry breaking events. Symmetry broken oscillations are observed in $N=3,5,10\text{ and }20$ CROW systems for higher input powers and detunings. Periodic switchings between different pairs of field intensities are also observed. Due to the access of higher input powers to all resonators, richer nonlinear phenomena are observed in CROW systems with all resonators coupled to input waveguides. Extended regions of $N$-level SSBs and chaotic oscillations are observed in this type of CROW configuration.

Future research will address the dynamic behavior of optical fields in the resonators and the effects of asymmetry in different parameters on the homogeneous states of the systems. SSBs in other complex arrangements of microring-resonators will also be addressed. The controllable distribution of light field intensities amongst different resonators could be a key feature for large-scale optical computing and light-field steering in integrated photonics. The combination of linear coupling between different resonators and optical nonlinearities in high-Q microresonators makes the CROW systems a promising candidate for integrated optical neural networks. CROW systems are also promising candidates for observing symmetry broken vector solitons with $N$ different values of circulating intensities~\cite{xu2021spontaneous}. These can be useful for generating $N$ distinct interconnected frequency combs which would be very useful in neuromorphic computing, telecommunications and especially in space technologies due to compactness. Together with the latest concepts of dispersion engineering~\cite{pal2023machine, li2020real, fujii2020dispersion} the studied nonlinear effects in this work will lead to a lot more interesting soliton dynamics.

\begin{acknowledgments}
This work was supported by the European Union’s H2020 ERC Starting Grant "CounterLight" 756966 and the Max Planck Society. AG and AP acknowledges the support from Max Planck School of Photonics. LH acknowledges funding provided by the SALTO funding scheme from the Max-Planck-Gesellschaft (MPG) and Centre national de la recherche scientifique (CNRS).\\
AG and  AP contributed equally to this work. PDH, AG and LH defined the research project. AG performed the theoretical analysis with support from LH and AP. AP, AG and LH completed the numerical simulations. AG, AP, LH and PDH wrote the manuscript with the help of all other authors. PDH and LH supervised the project.

\end{acknowledgments}

\bibliography{Refs}

\newpage
\onecolumngrid

\newpage

\appendix

\section{3 resonator CROW - Analytical Solutions}
\label{A}
For three resonator ``input to all" CROW system the equations of motion takes the form:
\begin{subequations}
\begin{eqnarray}
\frac{\partial \psi_1}{\partial \tau} &=& -\left(1 + i\zeta_1\right) \psi_1 + i j\psi_2 + i|\psi_1|^2\psi_1 + f,\\
\frac{\partial \psi_2}{\partial \tau} &=& -\left(1 + i\zeta_2\right) \psi_2 + i (j\psi_1 + j\psi_3)  + i|\psi_2|^2\psi_2 + f\\
\frac{\partial \psi_3}{\partial \tau} &=& -\left(1 + i\zeta_3\right) \psi_3 + i j\psi_2 + i|\psi_3|^2\psi_3 + f.
\end{eqnarray}
\label{appdxSSeqns}
\end{subequations}
Here $\zeta_i$ represents the normalized detuning of the $i^{\text{th}}$ resonator. In steady state, all the equations become equal to zero. Solving the steady state equations and considering equal input to all the resonators, one can obtain
\begin{subequations}
\begin{eqnarray}
\Psi_1^3 - \Psi_3^3 + 2\zeta_3\Psi_3^2 - 2\zeta_1\Psi_1^2 + (1 + \zeta_1^2)\Psi_1 - (1 + \zeta_3^2)\Psi_3 = 0,\\
\Psi_2^3 (C_3^2 + 1) + \Psi_2^2 (2C_2C_3 + 2C_4) + \Psi_2(C_2^2 + C_4^2) = \Psi_1 C_5,\\
\left|-\left(1 - \frac{j^2}{D_1}- \frac{j^2}{D_3}\right) + i(\Psi_2 - \zeta_2)\right|^2\Psi_2 = \left|1 - i\left(\frac{j}{D_1} + \frac{j}{D_3}\right)\right|^2 F,
\end{eqnarray}
\label{appdxSSsolnIACROW}
\end{subequations}
where $\Psi_n = |\psi_n|^2$, $D_n = -(1 + i\zeta_n) + i \Psi_n$ (for $n = 1,2$ or $3$), $C_1 = 1 - \zeta_2\zeta_3 - j\zeta_3$, $C_2 = C_1 + \zeta_2\Psi_3 + j\Psi_3$, $C_3 = \zeta_3 - \Psi_3$, $C_4 = -C_3  - \zeta_2 - j$, $C_5 = \{1 + C_3 (\Psi_1 - \zeta_1 - j) - j\zeta_1 + j\Psi_1\}^2 + \{\zeta_1 - \Psi_1 + C_3 + 2j\}^2$, and $F = |f|^2$.

For ``input to end" CROW system, $f_2 = 0$, therefore Eqs.~\eqref{appdxSSsolnIACROW} takes the form
\begin{subequations}
\begin{eqnarray}
\label{appdxSSsolnIECROWa}
\Psi_1^3 - \Psi_3^3 + 2\zeta_3\Psi_3^2 - 2\zeta_1\Psi_1^2 + (1 + \zeta_1^2)\Psi_1 - (1 + \zeta_3^2)\Psi_3 = 0,\\
\label{appdxSSsolnIECROWb}
\Psi_2^3 -2\zeta_2 \Psi_2^2  + \Psi_2(1 + \zeta_2^2) = \Psi_1 \left|j\left(\frac{D_3}{D_1}+1\right)\right|^2,\\
\label{appdxSSsolnIECROWc}
\left|D_2 + j^2\left(\frac{1}{D_1} + \frac{1}{D_3}\right)\right|^2\Psi_2 = \left|\left(\frac{j}{D_1} + \frac{j}{D_3}\right)\right|^2 F.
\end{eqnarray}
\label{appdxSSsolnIECROW}
\end{subequations}

\section{Field intensity crossings for $\mathbf{N=3}$ ``input to end" CROW systems}
\label{B}
For symmetric CROW system, Eq.~\eqref{appdxSSsolnIECROWb} takes the form,
\begin{eqnarray}
\Psi_2^3 -2\zeta \Psi_2^2  + \Psi_2(1 + \zeta^2) - 4j^2\Psi_1 = 0,
\label{SymappdxSSsolnIECROW}
\end{eqnarray}

where we have considered $\zeta_1=\zeta_2=\zeta_3=\zeta$ and $D_1=D_2=D_3$. At the crossing points, $\Psi_1 = \Psi_2$,
\begin{eqnarray}
\Psi_2 = \zeta \pm \sqrt{4j^2 - 1},
\label{SymappdxPsi2Cross}
\end{eqnarray}

Further using Eq.~\eqref{appdxSSsolnIECROWc}, at the crossing points, one can obtain,
\begin{eqnarray}
F = |f|^2 = \left|\frac{-3\pm i\sqrt{4j^2 - 1}}{2}\right|^2 \left(\zeta \pm \sqrt{4j^2 - 1}\right),
\label{SymappdxFCross}
\end{eqnarray}
The conditions for having different numbers of crossing are mentioned in the following table.
\begin{table}[h]
    \centering
    \begin{tabular}{c|c}
        \textbf{Number of crossing points}\; & \;\textbf{Conditions} \\[1ex] 
        \hline\\[-1ex] 
        $0$\; & $4j^2 < 1$ \\
        $1$\; & \; $4j^2 > 1$ and $\sqrt{4j^2 - 1}>\zeta$\\
        $2$\; & \; $4j^2 > 1$ and $\sqrt{4j^2 - 1}<\zeta$\\[-1ex] \;
    \end{tabular}
    \caption{Number of field intensity crossing points for $N=3$}
    \label{tab:my_label}
\end{table}

\begin{figure}[ht]
\includegraphics[width=0.7\columnwidth]{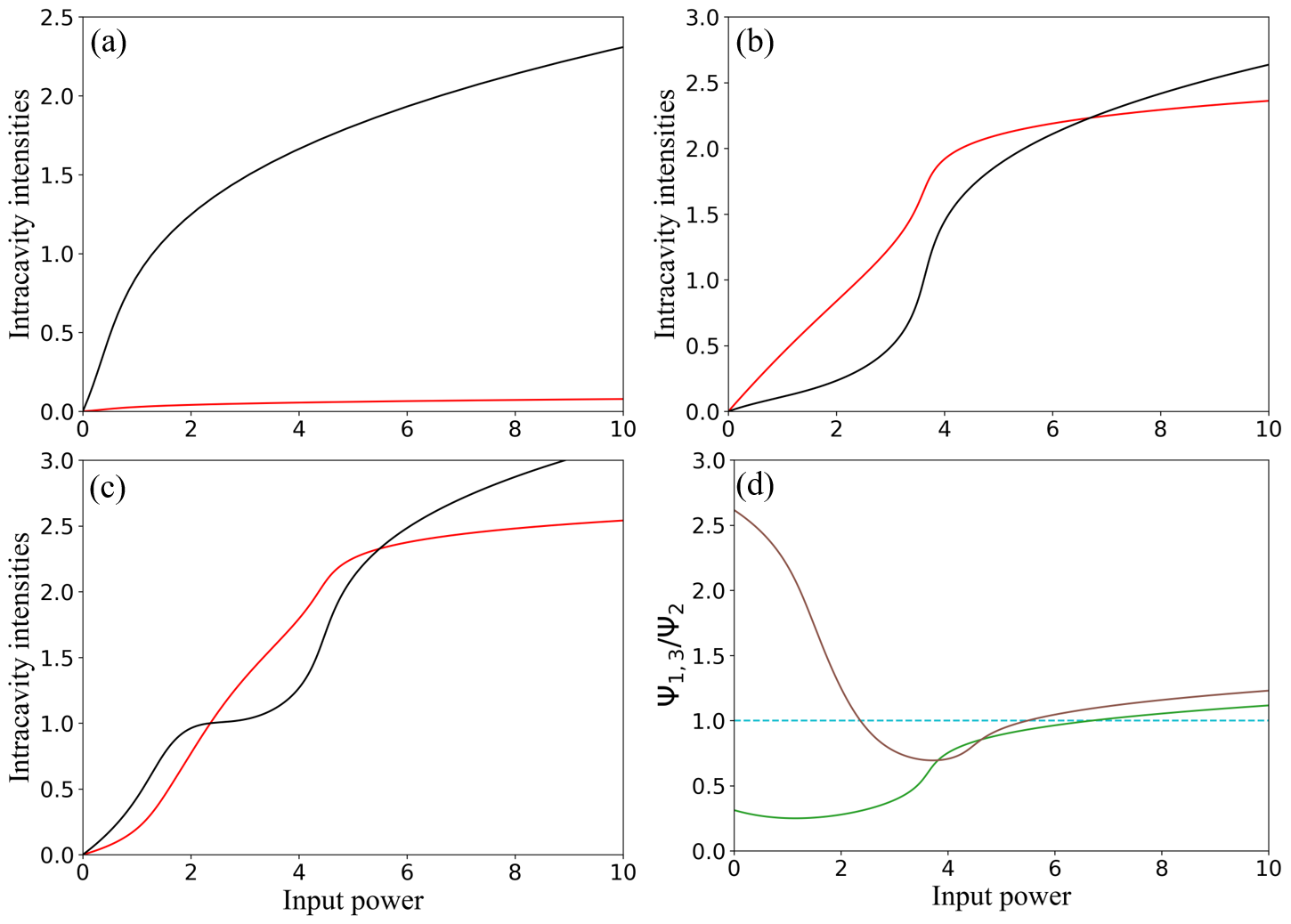}
\caption {\textit{Field intensity crossings for $N=3$ ``input to end" CROW systems}.
Panels (a-c) show input power scan of resonator light intensities for CROW system with $N = 3$. In panel (a) the end-resonator field  intensities (shown in black) do not cross the field  intensity in the middle resonator (shown in red). The middle resonator field intensity crosses end resonator field  intensities once in panel (b) and twice in panel (c). Panel (d) shows the ratio of field intensities circulating in end resonators and middle resonator. Brown line shows the ratio for (b), and green line shows the ratio for (c). Panel (d) demonstrates the potential for relative power distributions among the resonators in $N=3$ CROW systems. Used parameters: (a) $\zeta = 0.5$, $j = 0.1$, (b) $\zeta = 0.5$, $j = 1$, and (c) $\zeta = 1.66$, $j = 0.6$.
}
\label{Oscillations_sm_V2}
\end{figure}
\section{Eigen value analysis for $\mathbf{N=3}$ CROW system}
\label{C}
For $N=3$ CROW system, the normalized coupled LLE equations take the form:
\begin{subequations}
\begin{eqnarray}
\frac{\partial \psi_1}{\partial \tau} &= -\left(1 + i\zeta\right) \psi_1 + ij\psi_2 + i|\psi_1|^2\psi_1 + f,\\
\frac{\partial \psi_2}{\partial \tau} &= -\left(1 + i\zeta\right) \psi_2 + i\left(j\psi_1 + j\psi_3\right) + i|\psi_2|^2\psi_2,\\
\frac{\partial \psi_3}{\partial \tau} &= -\left(1 + i\zeta\right) \psi_3 + ij\psi_2 + i|\psi_3|^2\psi_3 + f,
\label{N3LLEquations}
\end{eqnarray}
\end{subequations}
where $\zeta$ is the normalized detuning, $j$ is the normalized coupling, $\psi_n$ is the normalized field envelop in the $i^{\text{th}}$ resonator and $f$ is the input field amplitude. If we define $\psi_n = \bar{\psi_n} + \delta \psi_n$, where $\bar{\psi_n}$ is the steady state value and $\delta \psi_n$ is the infinitesimal perturbation. Considering the complex conjugates of each field envelopes, the evaluation equations for the perturbations can be written as $\dot{A} = \mathcal{J}A$, where $A = \left[ \delta \psi_1, \delta \psi_1^*, \delta \psi_2, \delta \psi_2^*, \delta \psi_3, \delta \psi_3^* \right]^T$. The Jacobian matrix $\mathcal{J}$ can be written as

\begin{eqnarray}
\mathcal{J} = \begin{bmatrix} \Delta_1 & i|\bar{\psi_1}|^2 & ij & 0 & 0 & 0 \\ -i|\bar{\psi_1}|^2 & \Delta_1^* & 0 & -ij & 0 & 0 \\ ij & 0 & \Delta_2 & i|\bar{\psi_2}|^2 & ij & 0 \\0 & -ij & -i|\bar{\psi_2}|^2 & \Delta_2^* & 0 & -ij \\ 0 & 0 & ij & 0 & \Delta_3 & i|\bar{\psi_3}|^2 \\ 0 & 0 & 0 & -ij & -i|\bar{\psi_3}|^2 & \Delta_3^*,\end{bmatrix}
\label{Jacobian}
\end{eqnarray}
where $\Delta_n = -\left(1+i\zeta-i2|\bar{\psi_n}|^2\right)$. When the real part of any of the eigenvalues of the matrix becomes positive, the system becomes unstable to perturbations.
\clearpage
\section{Origin of the novel symmetry breaking mechanism in $\mathbf{N=3}$ CROW system}
\label{D}
\begin{figure}[ht]
\includegraphics[width=0.5\columnwidth]{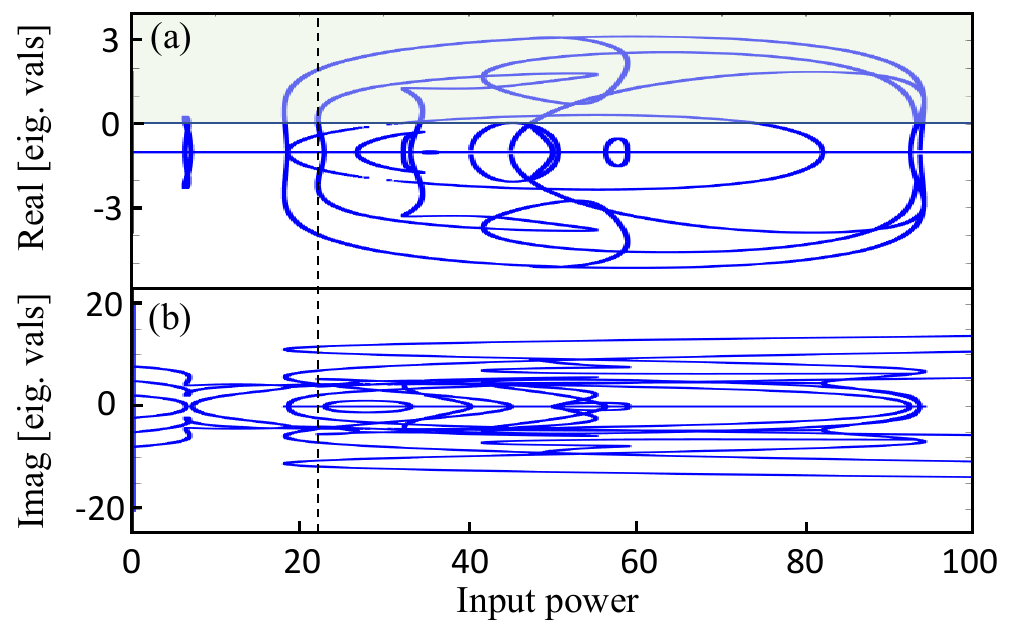}
\caption {\textit{SSB mechanism in $N=3$ CROW system}.
Panel (a) and (b) show real and imaginary parts of eigenvalues of the Jacobian matrix of the system as given by~\eqref{Jacobian}. The shaded region in panel (a) depicts the region where y-axis is above $0$. When the real part of any of the eigenvalues enters the shaded region, the system becomes unstable. The dashed line (which is at same place as the dashed line in Fig.~\ref{SSB_step}) depicts the input power value for which real part of one eigenvalue of the system goes above zero. The SSB between the circulating intensities in the end resonators occurs at this point.
}
\label{eig_vals_sm}
\end{figure}

In the main text, we have discussed about the occurrence of a novel type of SSB in $N=3$ CROW system, where the system jumps from an initial symmetric state to an isolated set of asymmetric solutions via oscillations. The system starts to oscillate when the real part of any of its' eigenvalues goes above zero. This is depicted in Fig.~\ref{eig_vals_sm}.

\clearpage
\section{Oscillations in CROW systems}
\label{E}
In the main paper, we have observed different types of oscillations in ``input to end" CROW systems. We have also observed the perfect periodic switching in $N = 3$ CROW systems. In this section, we will observe different types of oscillations present in the $N>3$ CROW systems with different input conditions.
\begin{figure}[ht]
\includegraphics[width=0.8\columnwidth]{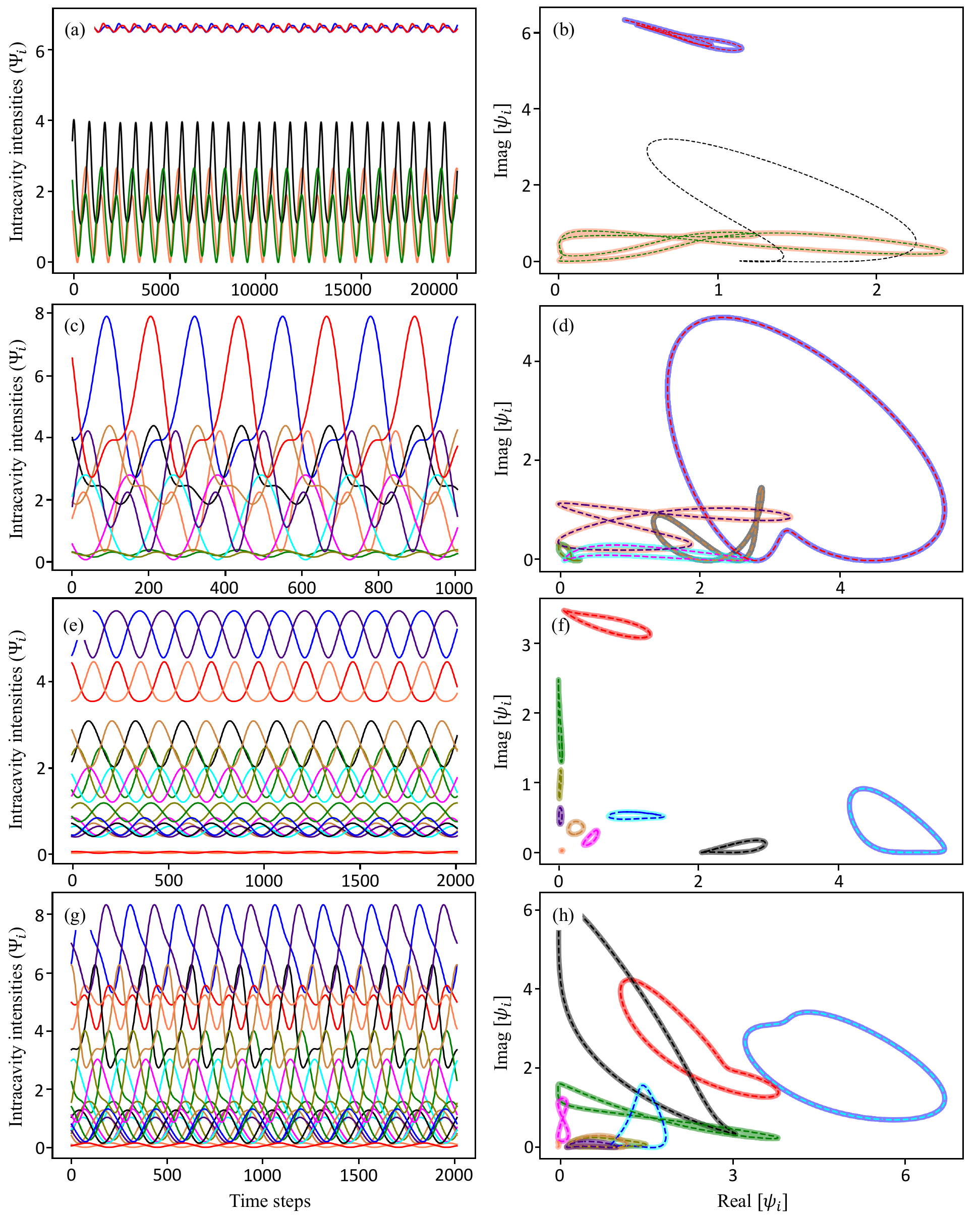}
\caption {\textit{Switchings in ``input to end" CROW systems}.
Periodic switchings of the fields within coupling-wise symmetric resonators for $N = 5$ (a-b), $N = 10$ (c-d), $N = 20$ (e-f,g-h) are depicted. Perfect sinusoidal switchings (a,c,e,g) are confirmed by the complete overlaps of the phase space plots (b,d,f,h). Used parameters: (a) $|f|^2 = 108.34$, $\zeta = 3$, $j = 1$, (a) $|f|^2 = 39.18$, $\zeta = 4.2$, $j = 3.5$, (a) $|f|^2 = 67.53$, $\zeta = 4$, $j = 3$, (a) $|f|^2 = 120.05$, $\zeta = 4$, $j = 3$. Time step for integration is $0.005$.
}
\label{Switching_sm}
\end{figure}

\begin{figure}[ht]
\includegraphics[width=0.8\columnwidth]{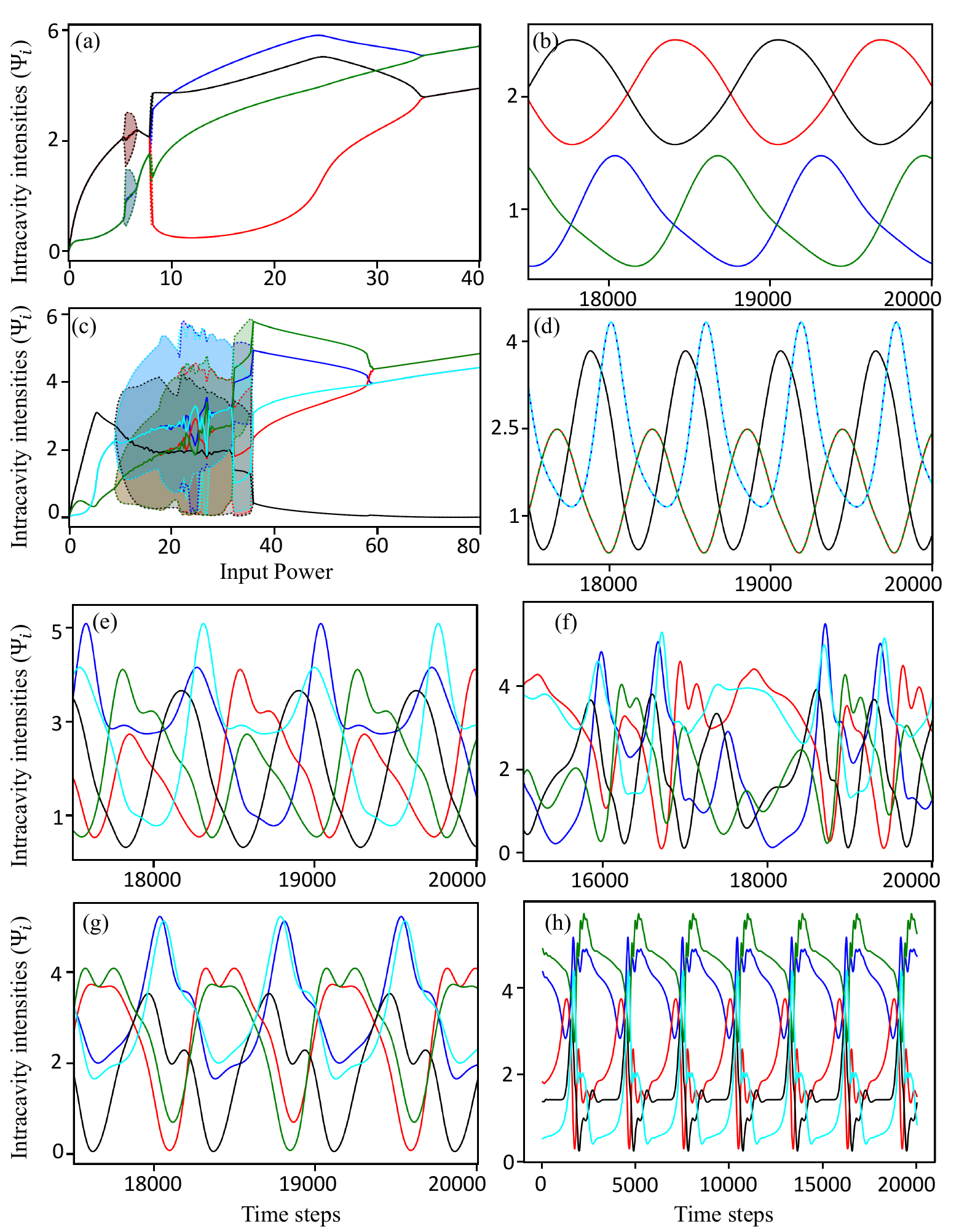}
\caption {\textit{Oscillations in ``input to all" CROW systems}.
(a) Input power scan of resonator light intensities for CROW system with $N = 4$, $\zeta = 1.5$ and $j = 1$. Evolutions of field intensities over time in the oscillatory region (shaded region in (a)) $|f|^2 = 5.68$. (c) Input power scan of resonator light intensities for CROW system with $N = 5$, $\zeta = 2,5$ and $j= 2$. (d) and (e) shows oscillations with maintained symmetry between fields within resonators with symmetric coupling conditions, whereas (f) shows switching oscillations between them. (g) depicts 5-field chaos. (h) shows near-switching oscillations. (i)-(j) shows completely asymmetric periodic oscillations. $|f|^2 = 12.93 \text{(d), }20.20 \text{(e), }24.44 \text{(f), }29.09 \text{(g), }34.14 \text{(h)}$. Time step for integration is $0.005$.
}
\label{Oscillations_sm_V2   }
\end{figure}

\end{document}